\documentclass[twoside,fleqn]{article}
\usepackage{psfig}
\usepackage{epsfig}
\usepackage{espcrc2}

\newcommand{\be}{\begin{equation}}
\newcommand{\ee}{\end{equation}}

 1
 1
 1

\title{How Should We Modify the High Energy Interaction Models ?}
\author{A.D.Erlykin\address{P.N.Lebedev Physical Institute - Moscow, Russia} 
\thanks{Corresponding author. Tel +7 095 1358737, fax +7 95 1357880, e-mail erlykin@sci.lebedev.ru},
A.W. Wolfendale\address{Department of Physics, University of
Durham - Durham, UK}}

\begin{document}

\begin{abstract}
An analysis has been made of the present situation with respect to 
the high energy
hadron-nucleus and nucleus-nucleus interaction models as applied to
cosmic rays. As is already known,
there are inconsistencies in the interpretation of 
experimental data on the primary mass composition, which appear when 
different EAS components are used for the analyses, even for the same
experiment. In the absence of obvious experimental defects, there is a
clear need for an improvement to the existing models; we argue that the most
promising way is to enlist two effects which should be present in
nucleus-nucleus collisions but have not been allowed for before. These are: 
a few percent energy transfer into the EAS electromagnetic component
due to electron-positron pair production or electromagnetic
radiation of the quark-gluon plasma and a small slow-down of the
cascading process in its initial stages associated with the extended
lifetime of excited nuclear fragments. The latter process displaces the
shower maximum deeper into the atmosphere.     
\end{abstract}
\maketitle

\section{Introduction}

One of the possible ways to assess the quality of the interaction
models is to use them for the determination of the primary mass
composition from the analysis of different cosmic ray components.
The ideal case is to get a consistent mass composition using the same
model and different components. As the variety and precision of
experimental data and the quality of the theoretical interaction
models are improved, more evidence appears that their quality is
still not good enough. The mass composition derived with the use of
electromagnetic and muon components is systematically lighter than when
hadrons and muons are used. Mass composition in the knee derived using
measurements of the depth of maximum is lighter than that from all
ground-based measurements \cite{Sword}. All these facts indicate that the
existing model needs further improvement. 

\section{The evidence from the ground-based measurements}

The variety of observed EAS components and precise measurements of
their characteristics by KASCADE experiment allowed this collaboration
to derive the primary mass composition using the multivariate analysis
of data. It was found that not only do different methods applied to
the same set of observables give different results, but that there is
a systematic difference between the results obtained using the same
method, but different components. In Figure 1, taken from \cite{Roth}, the
mean logarithmic mass $\langle lnA \rangle$ derived from different
sets of observables is shown vs. the so called {\em truncated} muon
number $N_\mu^{tr}$, which is an observable adopted as a measure of
the primary energy, independent of the primary mass. The fundamentsl
problem is that all the $\langle lnA \rangle$ values should be the
same at the same value of $N_\mu^{tr}$, and they are not. 

A number of features are of interest and, presumably, of some
importance: \\
(i) the use of the electron size $N_e$ results in an average lighter
composition (~Fig.1a~). On the other hand, omitting electrons and
using just muons and hadrons results in a heavier composition
(~Fig.1b~); \\
(ii) considering the mean values from the two sets of data (~filled-in
circles~), the ratio of $\langle \langle lnA \rangle \rangle$ for(b)
to (a) increases smoothly with $lgN_\mu^{tr}$, i.e. the discrepancy
between (b) and (a) rises with increasing primary energy (~Fig.1c~).
\begin{figure}[ht]
\psfig{figure=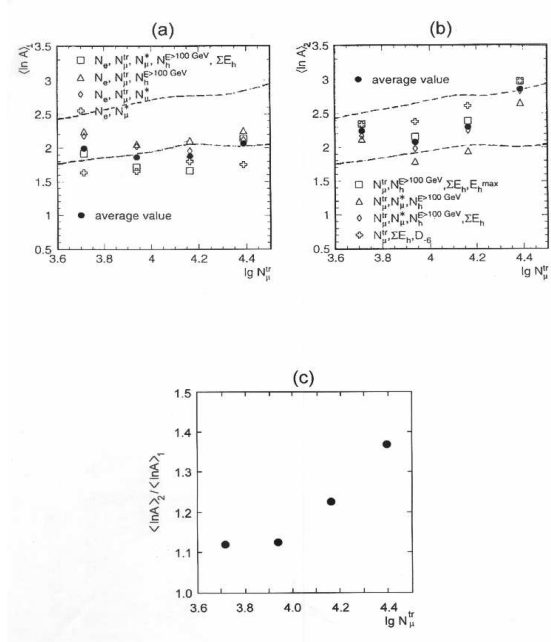,width=7.6cm, height=8.6cm}
\caption{\footnotesize Mean logarithmic mass $\langle lnA \rangle$
from the analysis \cite{Roth} of different sets of observables
vs. $lgN_\mu^{tr}$. The two dashed lines indicate alternative
predictions of the Single Source Model \cite{EW1}, with the upper line
being our latest and 'best estimate'. The sets displayed in (b) do not
include the observable $N_e$. It is seen that omitting $N_e$ results
in a heavier mass composition. (c) The ratio of mean values of
$\langle lnA \rangle$ from sets (b) and (a).}
\label{fig1}
\end{figure}
This difference points to inadequacies in the models used for the
analysis of the experimental data and provides an impetus to correct
them. However the corrections should not be radical, because the
difference between the $\langle lnA \rangle$ is not large, typically
$\delta \langle lnA \rangle \approx$ 0.2 - 0.4. The mentioned
systematic difference can be an indication that the energy
distribution between the different shower components is slightly
different from that in the models: specifically, the actual mean
number of electrons $N_e$ in EAS appears to be higher, that of muons
$N_\mu$ slightly lower and that of hadrons $N_h$ lower still than in
the models. 

It is well known from the present models that for the same primary
energy the number of muons in nuclei-induced showers is higher than in
proton-induced ones. On the other hand the number of electrons and
hadrons in nucleus-induced showers observed in the lower half of the
atmosphere is lower than in proton ones. If one has an opportunity to
measure the primary energy, by muons, Cherenkov light or another
technique, and finds that the shower has a low $N_\mu$ or a high $N_e$
(~actually the ratio $\frac{N_\mu}{N_e}$ is important~) the conclusion
will be that this shower is initiated by a proton or light nucleus. On
the contrary, if one finds that the shower has a low $N_h$ the
conclusion will be the opposite, i.e. that the shower is initiated by
a heavy nucleus. This is exactly what is observed in the showers at
sea level.  

The same conclusion can be drawn from an analysis of KASCADE event
rates \cite{Anton}. Both the muon and hadron trigger rates, observed by
KASCADE, are lower than  expected from the model calculations. This
discrepancy indicates again that the actual numbers of muons and
hadrons in EAS are lower than in the models, although the energy
region responsible for the trigger rates is lower than that analysed
for the mass composition around the knee. The analysis of this
discrepancy indicates that the needed reduction of the number of muons
in the model should be about 6\%, for hadrons it is bigger - about
29\%. Because muons and hadrons are the products of hadronic cascades,
it is evident from the energy balance that to reduce the energy
contained in the hadronic cascade one has to increase the energy
transferred into the electromagnetic cascade.

\section{Numerical estimates} 

\subsection{Balance between the EAS components}

To evaluate the effect of the proposed change of the balance between
different cascade components we applied the semi-quantitative analytical
approach, details of which can be found in \cite{EW2}. The results of the
calculation for our basic set of parameters are shown in Figure 2a by
a full line. According to our suggestion we increased the energy fraction
transferred by nucleons into the electromagnetic component
$K_\gamma^N$ from 0.20 to 0.26. The result is shown in Figure 2a by
the dashed line: the muon energy at sea level decreased by $\sim$ 6\%,
the hadron energy decreased by $\sim$ 23\%, the energy transferred
into the electromagnetic component increased by $\sim$ 2\%.
\begin{figure}[ht]
\psfig{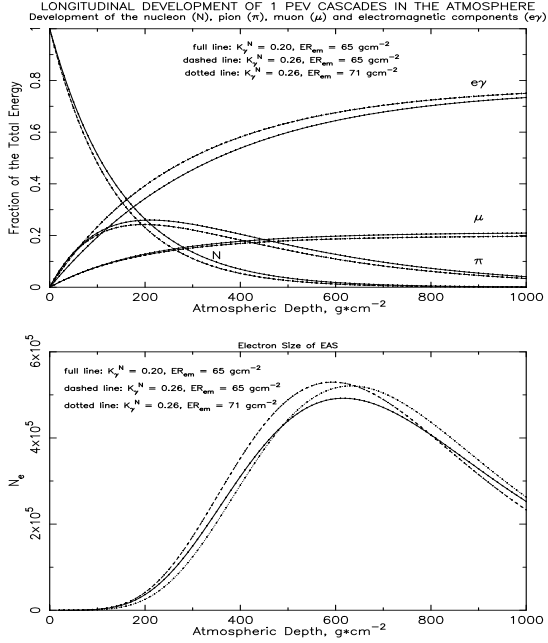}
\caption{\footnotesize The longitudinal development of 1 PeV cascades
in the atmosphere: (a) Fractions of the total energy carried by
nucleons ($N$), pions ($\pi$), muons($\mu$) and transferred into the
electromagnetic ($e\gamma$) component; (b) Electron size of the shower
$N_e$. Basic parameters: full line: $K_\gamma^N$ = 0.20, $ER$ = 65
gcm$^{-2}$; dashed line: $K_\gamma^N$ = 0.26, $ER$ = 65 gcm$^{-2}$;
dotted line: $K_\gamma^N$ = 0.26, $ER$ = 71 gcm$^{-2}$.}
\label{fig2}
\end{figure}
\subsection{The triangle diagrams}

The balance of the energy contained in the major EAS components is
convenient to analyse using the so-called 'triangle diagrams'
\cite{Danil}. If the height of an equilateral triangle is equal to 1,
then for each point inside this triangle the sum of the distances to
its sides is equal to 1. If we know the energy fractions carried by
the electromagnetic ($\delta_{e\gamma}$), muon ($\delta_\mu$) and
hadron ($\delta_h$) components at the observation level, so that
$\delta_{e\gamma} + \delta_\mu + \delta_h = 1$, then each shower can be
presented by a single point inside the triangle. Our basic shower
(~Fig.2, full line~) is shown in Fugure 3 by a full circle. The
desired direction for the shift of the energy balance in the modified
model is shown by the straight line arrow in Figure 3b. 
\begin{figure}[ht]
\psfig{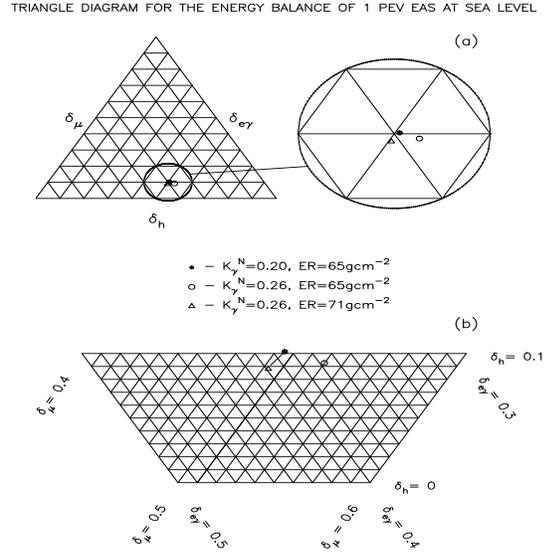}
\caption{\footnotesize Triangle diagram for a 1 PeV shower at sea
level. (a) Large scale diagram with the inset magnifying the indicated
region. (b) Small scale part of the full diagram. The straight arrow
to the left of the full circle indicates the desired modification of
the EAS energy balance.}
\label{fig3}
\end{figure}

However, despite the increase of the energy {\em transferred} into the
electromagnetic component, the {\em preserved} electromagnetic energy
and the electron size of the shower at sea level decreased by 14\% due
to its faster development and then faster attenuation of the cascade 
(~Figure 2b~). The point in the triangle diagram moved in the
different direction (~open circle~). The similar result is obtained if
we increase the total
inelasticity $K_{tot}^N$ with $K_\gamma^N =
\frac{1}{3}K_{tot}^N$. Therefore {\em the mere increase of the mean
value of $K_\gamma^N$ cannot give the required result.}    

We argue that another effect that {\em should} be present will bring
about the desired effect: the slowing down of the development of the
cascade in its initial stages. For illustration purposes we slow down
the development of the hadronic and electromagnetic cascade by
increasing the elongation rate {\em ER} from 65 gcm$^{-2}$ to 71
gcm$^{-2}$, preserving $K_\gamma^N$ = 0.26. The result is shown in
Figure 2 by dotted lines. The direction and the magnitude of the
changes is now correct. Thus we conclude that {\em the increase of the
energy transferred into the electromagnetic component combined with
the slowing down of the development of cascades in their initial
stages is the most realistic way to improve the particle interaction
model and to achieve a consistent estimate of the primary mass composition.}  

\section{Theoretical arguments}

Besides all these arguments, which are purely phenomenological, there
are also theoretical arguments which lend support to the phenomenological
consideration. Nearly all of them are related to processes which
appear in nucleus-nucleus (~AA~) interactions. Their details are given
in \cite{EW2} and here we just enumerate them.

In AA collisions with a small impact parameters (~central collisions~)
one has to expect the production of $e^+e^-$-pairs including the
multiple pair production. This process provides an additional energy
transfer into the electromagnetic component and it was not taken into
account in
the present models. An additional energy transfer into the
electromagnetic component can arise also from an excess of direct
photons, which has been predicted theoretically as a signature of the
quark-gluon plasma and is now observed in AA-collisions both at low
and high transverse momentum.

In AA collisions with a large impact parameters (~peripheral
collisions~) a projectile nucleus fragments into few pieces of
different mass. Some of them are excited and after de-excitation give
rise to MeV gamma-ray lines, observed from 'discrete' sources and the
interstellar medium. The lifetime of the excited fragments varies from
a 'nuclear' time $\sim 10^{-23}$ sec to millions of years. 
For AA-interactions at PeV energies both the
lifetime before the de-excitation and the energy of emitted
gamma-quanta are extended by the factor of 10$^5$ - 10$^6$ due to
relativistic effects. As a consequence one can expect an additional
sub-PeV electromagnetic cascade to be initiated a few hundred meters
below the point of the first interaction. This effect will slow down
the development of the electromagnetic cascade and shift its maximum.

\section{Conclusions}

The inconsistencies in the interpretation of the experimental data on
the primary mass composition, obtained when different EAS components
are used for the analysis, indicate the need for some improvement to
the models which were used hitherto. We propose that the most
promising way is to introduce an additional (~a few percent~) energy
transfer into the EAS electromagnetic component combined with a
slowing down of the cascade development in its initial stages, which
is followed by a small (~20-30 gcm$^{-2}$~) shift of the shower
maximum into the deeper atmosphere. The most likely processes which
can be responsible for such changes are those which occur in
AA-collisions and they should indeed be present {\em at some level}.
The importance of these processes is expected to grow with energy and
offers the hope of resolving some controversies at very high energies.


\begin{thebibliography}{99}
\bibitem{Sword} Swordy S.P. et al., 2002, astro-ph/0202159 
\bibitem{Roth}  Roth M. et al., 2001, Proc. 27th ICRC, Hamburg, {\bf 1}, 88 
\bibitem{EW1} Erlykin A.D., Wolfendale A.W. 1998, Astropart. Phys.,
{\bf 9}, 213
\bibitem{Anton} Antoni T. et al., 2001, J. Phys. G: Nucl. Part. Phys.,
{\bf 27}, 1785
\bibitem{EW2} Erlykin A.D., Wolfendale A.W. 2002, Astropart. Phys.,
{\bf 18}, 151
\bibitem{Danil} Danilova T.V., Erlykin A.D., 1983, Proc. 18th ICRC,
Bangalore, {\bf 5}, 262
\end{thebibliography}
\end{document}